\documentclass[conference]{IEEEtran}
\IEEEoverridecommandlockouts

\usepackage{graphicx}
\usepackage{amsmath}
\usepackage{amssymb}
\usepackage[caption=false]{subfig}
\usepackage[noadjust]{cite}
\usepackage{float}
\usepackage{color}
\usepackage{url}

\definecolor{armygreen}{rgb}{0.0, 0.5, 0.0}

\DeclareMathOperator*{\argmax}{arg\,max}
\DeclareMathOperator*{\argmin}{arg\,min}

\graphicspath{{img/}}

\begin{document}

\title{mmWave Channel Estimation via Approximate Message Passing with Side Information
\thanks{Rush was supported in part by NSF CCF $\#1849883$.
Baron was supported in part by NSF ECCS $\#1611112$.
}}

\author{\IEEEauthorblockN{Dror Baron}
\IEEEauthorblockA{\textit{Dept. Electrical \& Computer Eng.} \\
\textit{North Carolina State University}\\
Raleigh, NC, USA \\
barondror@ncsu.edu}
\and
\IEEEauthorblockN{Cynthia Rush}
\IEEEauthorblockA{\textit{Department of Statistics} \\
\textit{Columbia University}\\
New York, NY, USA \\
cynthia.rush@columbia.edu}
\and
\IEEEauthorblockN{Yavuz Yapici}
\IEEEauthorblockA{\textit{Dept. Electrical \& Computer Eng.} \\
\textit{North Carolina State University}\\
Raleigh, NC, USA \\
yyapici@ncsu.edu}
}

\maketitle

\begin{abstract}
This work considers millimeter-wave channel estimation in 
a
setting where 
parameters of the underlying mmWave channels are varying dynamically over time 
{and there is a single drifting path}.  In this setting, channel estimates at 
time block 
$k$ can be used as side information (SI)
when estimating the channel at block 
$k+1$. To estimate channel parameters, we employ 
an SI-aided 
(complex) approximate message passing algorithm and compare its performance to 
a benchmark based on orthogonal matching pursuit.
\end{abstract}


\section{Introduction}\label{sec:intro}

Mobile user demand for wireless data services has been increasing dramatically in recent years.
As the conventional sub-6 GHz communications spectrum is packed with existing wireless services, 
the \emph{millimeter-wave} (mmWave) frequency band has become a key asset for next-generation cellular networks. 
Along with increasing antenna array sizes at both sides of the transceiver, \emph{compressed sensing} (CS) 
based algorithms have received great attention in estimating mmWave channels. 
Owing to the mobility of users and scattering obstacles (moving cars and so on) in the communication environment, 
parameters underlying mmWave channels 
vary 
dynamically over time. These variations can either be estimated from scratch,
likely at the expense of 
significant 
training overhead, or tracked by making use of dynamic channel characteristics. The focus of our work
is to  
perform channel estimation using {\em approximate message passing} 
(AMP) aided by \emph{side information} (SI).
Our AMP-SI approach to channel estimation utilizes the dynamic channel structure,
leading to improved estimation quality and reducing training overhead.

{\bf Approximate Message Passing.} 
We use a class of low-complexity algorithms, referred to as 
AMP~\cite{DonMalMont09, krz12, MontChap11, rangan2019vector}, 
for channel estimation. 
AMP was originally introduced 
in the context of CS~\cite{DonohoCS,CandesRUP},
where one wishes to recover an unknown sparse vector $\boldsymbol{\beta}$ from 
noisy linear measurements $\mathbf{y}$ modeled as 
\begin{equation}
\mathbf{y} = \mathbf{A} \boldsymbol{\beta} + \mathbf{n},
\label{eq:CS}
\end{equation} 
where $\mathbf{A}$ is a measurement matrix with more columns than rows, and $\mathbf{n}$ is
{\em independent and identically distributed} (i.i.d.) noise. AMP 
iteratively estimates
$\boldsymbol{\beta}$ using a possibly non-linear denoiser 
function tailored to prior knowledge about $\boldsymbol{\beta}$. 
One key property of AMP is that under some technical conditions on the measurement matrix $\mathbf{A}$ and 
signal $\boldsymbol{\beta}$, observations at each iteration of the algorithm, referred to as \emph{pseudo-data}, 
are asymptotically (in the large system limit) equal in distribution to $\boldsymbol{\beta}$ plus 
i.i.d.\ Gaussian noise.

{\bf AMP with Side Information (AMP-SI).} Recently~\cite{Ma2018, Liu2019ISIT},  AMP-SI was introduced as an
algorithmic framework that incorporates
SI into AMP for CS tasks 
(\ref{eq:CS}).   
AMP-SI has been empirically demonstrated to have good reconstruction quality, and is easy to use. For example, we have proposed to use AMP-SI for a toy model for channel estimation in emerging 
mmWave 
communication systems~\cite{Rappaport2016mmW}, 
where the time dynamics of the channel structure allow previous channel estimates to be used as SI when estimating the current channel structure~\cite{Ma2018}. In Liu et al.~\cite{Liu2019ISIT}, the nice empirical performance of AMP-SI 
was strengthened through a rigorous performance analysis.  
For these reasons, it is not surprising that our novel approach to channel estimation outperforms a benchmark based on {\em orthogonal matching pursuit} 
(OMP)~\cite{Pati1993}
as evidenced in Sec.~\ref{sec:results}, 
and it is unlikely that other non-AMP based approaches would yield further improvements.

{\bf Contributions.} 
Our main insight in this paper is that the channel matrix can be 
represented sparsely over the domain of angles of arrival and departure.
This insight leads us to develop a denoiser within AMP-SI that monitors and
estimates paths with continuous angles of arrival and departure.
We use 2D spectral estimation within AMP-SI 
for a simplified problem with a single drifting path,
and will 
address increasingly complicated (and thus realistic) models. 

{\bf Notation.} Let $(\cdot)^*$ and $(\cdot)^{\rm H}$ denote the complex conjugate and Hermitian operations, respectively. 
We use $\textbf{0}_M$ and $\textbf{I}_M$ to represent a zero vector of size $M{\times}1$ and identity matrix of size $M{\times}M$, respectively. Next, $[\textbf{x}]_i$ stands for the $i$-th element of the vector $\textbf{x}$ and $[\textbf{M}]_{ij}$ for the $(i,j)$-th element of the matrix $\textbf{M}$.
A complex Gaussian distribution with mean $\textbf{m}$ and covariance $\textbf{C}$ is denoted by 
$\mathcal{CN}(\textbf{m},\textbf{C})$, 
and $\mathcal{U}[a,b]$ stands for the uniform distribution taking values between $a$ and $b$. 
Finally, 
the set of integers $\{1, 2, \ldots, M\}$ is denoted by $[M]$, and
the Dirac delta function, $\delta_{k\ell}$, takes the value $1$ if $k \,{=}\, \ell$, and $0$ otherwise.

\section{System Model} \label{sec:system}

\begin{figure}[!t]
\centering
\includegraphics[width=0.5\textwidth]{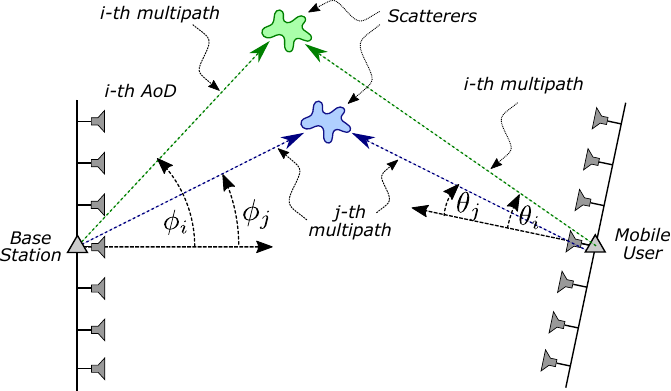}
\caption{
System model of mmWave communication.
Representative $i$-th and $j$-th multipaths are shown, 
along with corresponding angles of arrival 
$\{\theta_i,\theta_j\}$ and departure $\{\phi_i,\phi_j\}$.}
\label{fig:scenario}
\vspace{0.in}
\end{figure}

Consider point-to-point downlink communication in mmWave frequency spectrum
(Fig.\ \ref{fig:scenario}), where a {\em base station} (BS) communicates with a mobile {\em user equipment}
(UE). The number of
{\em transmit} and {\em receive} antennas at the BS and UE are $M_\mathsf{t}$ and $M_\mathsf{r}$, 
respectively, both of which form a {\em {\color{black}one}-dimensional} ({\color{black}1}D) {\em uniform linear array} 
(ULA). 

We study a blockwise transmission strategy that relies on block fading.
The downlink channel in the $k$-th transmission block, denoted by $\textbf{H}_k$, does not change during the $k$-th transmission block;
the channel matrix takes a new value in the next transmission block. We will assume that across blocks~$k$, 
there is some dependence or structure,
so that an estimate of the block $\textbf{H}_k$ can be used as {\em side information} (SI) 
in estimating $\textbf{H}_{k+1}$.  The relationship between $\textbf{H}_k$ and $\textbf{H}_{k+1}$ is detailed in Sec.~\ref{sec:time_var_model}, 
and the channel model in Sec.~\ref{sec:channel_model}.  
Our goal will be to estimate $\textbf{H}_k$ for each block $k$ from received signals, 
with the signal model specified in Sec.~\ref{sec:signal_model}.

\subsection{Transmission and Channel Models} \label{sec:channel_model}
 
In our transmission strategy,
the BS transmits {\em pilot} symbols (known to the UE) during the first $T_\mathsf{p}$ time slots of each transmission block, which consists of $T$ time slots in total. The UE estimates the channel using these  $T_\mathsf{p}$ pilot symbols together with the SI, 
which is obtained from the channel during previous transmission blocks. 
The BS uses the remaining $T{-}T_\mathsf{p}$ time slots to transmit 
payload data
(unknown to the UE), which is decoded by the UE using the channel estimate of the current transmission block.   

The respective downlink channel, $\textbf{H}_k$, within the $k$-th transmission block is given as follows,
\begin{align} \label{eq:channel}
\textbf{H}_k \,{=}\, \sqrt{ \frac{ M_\mathsf{t}M_\mathsf{r} }{ L_\mathsf{p} } } \sum\limits_{\ell=1}^{L_\mathsf{p}} \sqrt{\gamma} \alpha_{k,\ell} \, \textbf{a}(\theta_{k,\ell},M_\mathsf{r}) \textbf{a}(\phi_{k,\ell},M_\mathsf{t})^{\rm H},
\end{align} 
where $L_\mathsf{p}$ 
is the number of multipath components,
$\gamma$ is the \emph{signal-to-noise ratio} (SNR),
$\alpha_{k,\ell}$ is the complex gain of the $\ell$-th multipath component assumed to be distributed as 
$\mathcal{CN}(0,1)$ with uncorrelated gains for different paths, i.e.\
$ \mathbb{E}[\boldsymbol{\alpha}_{k} (\boldsymbol{\alpha}_{k}^*)^{\rm T}] =\sigma^2 \textbf{I}_{L_\mathsf{p}}$
where $\boldsymbol{\alpha}_k \,{=}\, [ \alpha_{k,1} \, \alpha_{k,2} \dots \alpha_{k,L_\mathsf{p}} ]^{\rm T}$. 
In addition, $\theta_{k,\ell}$ and $\phi_{k,\ell}$ are the \emph{angle-of-arrival} (AoA) and \emph{angle-of-departure} (AoD), respectively,
of the $\ell$-th multipath component with $\theta_{k,\ell} \sim \mathcal{U}[0,2\pi]$ and $\phi_{k,\ell} \sim \mathcal{U}[0,2\pi]$. 
Note that the distributions for the complex path gain and AoA/AoD folow from mmWave channel measurement studies~\cite{Rappaport2016mmW}. 
Furthermore, $\textbf{a}(\theta_{k,\ell},M_\mathsf{r})$ and $\textbf{a}(\phi_{k,\ell},M_\mathsf{t})$ 
represent the array steering vectors at the receive (UE) and transmit (BS) sides, 
respectively, where the $m$-th element of a generic array steering vector $\textbf{a}(\varphi,M)$ is
\begin{align}
\Big[ \textbf{a}(\varphi,M) \Big]_m =  \frac{1}{\sqrt{M}} {\rm exp} \left\lbrace j2\pi \frac{d_\mathsf{a}}{\lambda} \left( m{-}1\right) \sin\left( \varphi \right) \right\rbrace,
\end{align}
for $m \in [M]$, with an arbitrary phase $\varphi$ representing AoA/AoD, and $M$ elements 
representing the number of transmit/receive antenna elements,   
where $d_\mathsf{a}$ is the antenna element spacing of the ULA, and $\lambda$ is the carrier frequency wavelength.

Recall that our goal is to estimate $\textbf{H}_k$ for each block $k$ from received signals
and SI.
To this end, we will assume throughout that in our definition for $\textbf{H}_k$ \eqref{eq:channel} 
the scalar values $M_\mathsf{t}$, $M_\mathsf{r}$, $L_\mathsf{p}$, and $\gamma$ 
are known, and that $d_\mathsf{a}$ and $\lambda$ are also available to the ULA antenna array. 
Note that the number of paths $L_\mathsf{p}$ and SNR $\gamma$ 
are relatively stationary for point-to-point communications, and can be separately obtained 
over many transmission blocks. As a result, estimating the channel block $\textbf{H}_k$ boils down to 
estimating the $L_\mathsf{p}$ complex Gaussian 
{\em random variables} (RVs) 
$\alpha_{k, \ell}$ and the $2L_\mathsf{p}$ uniform RVs $\theta_{k, \ell}$ and $\phi_{k, \ell}$, from which the arrays $\textbf{a}(\theta_{k,\ell},M_\mathsf{r})$ and $\textbf{a}(\phi_{k,\ell},M_\mathsf{r})$ and thus 
$\textbf{H}_k$ can be recovered.

\subsection{Time Variation for Channel Parameters} \label{sec:time_var_model}

We now provide a model for the dynamic
relationship between $\textbf{H}_k$ and $\textbf{H}_{k+1}$. 
Our model was discussed by numerous authors
(c.f.,~\cite{Guo2016TraAng, Heath2016BeaTra, Kim2017RobBea} and references therein).
The complex path gain varies from one transmission block to another following a first order {\em auto regressive} (AR) 
process,
\begin{align}\label{eqn:time_evol_path}
\boldsymbol{\alpha}_{k{+}1} = \rho \boldsymbol{\alpha}_{k} + \textbf{u}^{\alpha}_k,   
\end{align} 
where $\rho$ is the correlation coefficient, and $\textbf{u}^{\alpha}_k$ is innovation 
with $\textbf{u}^{\alpha}_k \sim \mathcal{CN}( \textbf{0}_{L_\mathsf{p}}, \left(1{-}\rho^2\right) \textbf{I}_{L_\mathsf{p}})$. Note that \eqref{eqn:time_evol_path} represents 
variation in the complex path gains (i.e., small-scale fading) with correlation to the previous transmission blocks through $\rho$.   
Moreover, both AoA and AoD follow Gaussian 
innovation processes in the next transmission block,
\begin{align}
\boldsymbol{\theta}_{k{+}1} = \boldsymbol{\theta}_{k} + \textbf{u}^{\theta}_k \qquad \text{ and } \qquad
\boldsymbol{\phi}_{k{+}1} = \boldsymbol{\phi}_{k} + \textbf{u}^{\phi}_k, \label{eqn:time_evol_aod}
\end{align} 
where 
$\boldsymbol{\theta}_k \,{=}\, \left[ \theta_{k,1} \, \theta_{k,2} \dots \theta_{k,L_\mathsf{p}} \right]^{\rm T}$ 
and 
$\boldsymbol{\phi}_k \,{=}\, \left[ \phi_{k,1} \,  
\dots \phi_{k,L_\mathsf{p}} \right]^{\rm T}$
are the aggregate AoA and AoD vectors, respectively, 
and $\textbf{u}^{\theta}_k \sim \mathcal{N}( \textbf{0}_{L_\mathsf{p}},\sigma^2_{\theta}\textbf{I}_{L_\mathsf{p}})$ and $\textbf{u}^{\phi}_k \sim \mathcal{N}( \textbf{0}_{L_\mathsf{p}},\sigma^2_{\phi}\textbf{I}_{L_\mathsf{p}})$ are corresponding innovation vectors.

\section{mmWave Channel Estimation via AMP-SI} \label{sec:signal_model}

In this section, we consider mmWave channel estimation for the scenario described in Sec.~\ref{sec:system}, 
and explain how SI from previous transmission blocks enhances the estimation quality. To this end, we define an $M_\mathsf{t}{\times}1$ 
complex-valued unit-energy vector $\textbf{s}_{k,i}$, which represents the pilot symbol transmitted during the $i$-th time slot 
within transmission block~$k$. We also assume that $\textbf{s}_{k,i}$ is selected from an uncorrelated dictionary, i.e., 
$\mathbb{E}[\textbf{s}_{k,i} \textbf{s}_{k',i'}^{\rm H}] \,{=}\, \delta_{k k'} \delta_{i i'}$
where $k, k', i, i' \in [M_\mathsf{t}]$.

{\bf From signal estimation to matrix estimation.}
The received signal vector at the UE is given by
\begin{align}
    \textbf{y}_{k,i} = \textbf{H}_k \textbf{s}_{k,i} + \textbf{n}_{k,i},
\end{align}
where $\textbf{n}_{k,i}$ is measurement noise that follows $\mathcal{CN}(\textbf{0}_{M_\mathsf{r}},\textbf{I}_{M_\mathsf{r}})$. 
Considering pilot transmissions over $T_\mathsf{p}$ time slots, the aggregate received signal 
during the $k$-th transmission block, $\textbf{Y}_{k} \,{=}\, [\textbf{y}_{k,1} | \cdots | \textbf{y}_{k,T_\mathsf{p}}] \in 
{\mathbb{C}}^{ M_\mathsf{r} \times T_\mathsf{p}}$, 
where $[\textbf{v}_1 | \cdots | \textbf{v}_{T_\mathsf{p}}]$ denotes a matrix obtained by concatenating the column vectors $\textbf{v}_1, \ldots, \textbf{v}_{T_\mathsf{p}}$, is given by
\begin{align}\label{eq:received_signal}
    \textbf{Y}_{k} = \textbf{H}_k \textbf{S}_{k} + \textbf{N}_{k}.
\end{align} 
Note that $\textbf{S}_{k} \,{=}\, [\textbf{s}_{k,1} | \cdots  | \textbf{s}_{k,T_\mathsf{p}}]$ and $\textbf{N}_{k} \,{=}\, [\textbf{n}_{k,1} | \cdots |\textbf{n}_{k,T_\mathsf{p}}]$ are  
$M_\mathsf{t} \times T_\mathsf{p}$ and 
$M_\mathsf{r} \times T_\mathsf{p}$, respectively. Our goal is to estimate $\textbf{H}_k$ using \eqref{eq:received_signal} given 
observations $\textbf{Y}_{k}$ and pilot symbols $\textbf{S}_{k}$ and SI from previous transmission blocks.

As mentioned in Sec.~\ref{sec:intro},
we use AMP-SI to estimate the channel $\textbf{H}_k$ from the model~\eqref{eq:received_signal}.  
While~\eqref{eq:received_signal} is not identical to the CS
problem~\eqref{eq:CS}, taking the transpose of~\eqref{eq:received_signal} we have 
\begin{align}\label{eq:received_signal_transpose}
    \textbf{Y}_{k}^{\rm T} = \textbf{S}_{k}^{\rm T}\textbf{H}_k^{\rm T} + \textbf{N}_{k}^{\rm T},
\end{align}
which is more aligned with the AMP framework. In particular, we could 
modify (\ref{eq:received_signal_transpose})
by vectorizing $\textbf{Y}_{k}^{\rm T}$ and $\textbf{H}_k^{\rm T}$ and composing a measurement matrix $\mathbf{A}$ having $ \textbf{S}_{k}^{\rm T}$ 
repeated on the diagonal. However, this modification is not necessary, because AMP provides favorable results even when applied to 
multi-dimensional signals~\cite{Tan_CompressiveImage2014, Metzler2016_IEEE}. 
Therefore, we run AMP directly on the multi-dimensional problem.

{\bf AMP-SI with 2D denoisers.}
Consider a fixed block $k$.  We run AMP-SI on the matrix $\textbf{H}_k^{\rm T}$ directly, employing a well-chosen 2D (matrix) denoiser.  
Each AMP iteration will have access to pseudo-data
that is asymptotically (in the large system limit) equal in distribution to $\textbf{H}_k^{\rm T}$ plus a matrix of i.i.d.\ Gaussian noise, 
where the existing AMP theory allows us to
calculate a good approximation for the noise variance.  Importantly, the 2D denoiser we propose incorporates SI from the previous estimate at block $k-1$, where this SI is our estimate for the multipath parameters in block $k-1$, 
and our knowledge of the dynamics of these parameters per~\eqref{eqn:time_evol_path}-\eqref{eqn:time_evol_aod}.

Denoising a 2D matrix within AMP, as opposed to a vector, is non-standard.
That said, some related art (including by the authors) is encouraging.
For example, previous applications of AMP with multi-dimensional denoisers have provided encouraging empirical results in image 
reconstruction~\cite{Tan_CompressiveImage2014, Metzler2016_IEEE}. 
Beyond empirical results, a rigorous analysis by Ma \emph{et al.}~\cite{Ma_nonseparable19} 
provides performance guarantees for a family of multi-dimensional sliding window denoisers used within AMP.

{\bf Spectral estimation in 2D.}
Our proposed denoiser resembles work by Hamzehei and Duarte~\cite{Hamzehei2015compressive,Hamzehei2016compressive},
who performed {\em analog denoising} of 1D vectors within AMP in the context of spectral estimation.
Our denoisers resemble theirs, 
except that we perform 2D instead of 1D spectral estimation, 
with improved performance owing to SI from block~$k-1$.  
{\em Our main insight is that $H_k$
is sparse over the \underline{continuous} (AoA, AoD) domain.}
This insight leads us to develop a denoiser that monitors and
estimates paths with continuous $\phi$ and $\theta$.
Our work seems most related to Bellili et al.~\cite{Bellili2019generalized},
where the authors sparsify a linear inverse problem using Fourier arguments.
Our approach expands over theirs by performing 2D continuous spectral estimation within AMP.

To make the details of our denoiser tractable (Sec.~\ref{sec:one_path}), 
we begin with a simplified setting comprised of 
a {\em single drifting path}. That is, we assume $L_{\textsf{p}} = 1$ in channel model~\eqref{eq:channel}.  
While this paper introduces AMP-SI and our 2D spectral estimation denoiser for this simplified version of the problem, 
we aim to leverage these results and develop a series of denoisers addressing increasingly complicated (and thus realistic) 
models. 
Our current work will be extended to multiple paths with birth-death-drift dynamics 
between blocks~\cite{Ma2018}.

\section{One drifting path} 
\label{sec:one_path}

As mentioned previously, we set $L_{\textsf{p}} = 1$ in \eqref{eq:channel}, 
and model the channel in the $k$-th transmission block, $\textbf{H}_k$, as
\begin{align} \label{eq:channel_one_path}
\textbf{H}_k = \sqrt{ M_\mathsf{t}M_\mathsf{r} \gamma}\, \alpha_{k} \, \textbf{a}(\theta_{k},M_\mathsf{r}) \textbf{a}(\phi_{k},M_\mathsf{t})^{\rm H}.
\end{align} 
In this section, we introduce an AMP-SI algorithm for completing this parameter estimation task, 
and discuss some implementation details.

\textbf{AMP-SI details.}
The AMP-SI algorithm for estimating $\textbf{H}_k$ in \eqref{eq:channel_one_path} takes the following form.  Initialize 
the matrix estimate with $\widehat{\textbf{H}}_k^t = \mathbf{0}$, a zeros matrix, and 
at iteration $t \geq 0$, compute
\begin{equation}
\begin{split}
\textbf{R}^t &= \textbf{Y}_{k}^{\rm T} - \textbf{S}_{k}^{\rm T} \widehat{\textbf{H}}_k^t + \textbf{R}^{t-1} \langle div \,\, \eta_t(\mathbf{V}^t, \text{SI}_{k-1}) \rangle, \\
\widehat{\textbf{H}}_k^{t+1}&= \eta_t(\mathbf{V}^t, \text{SI}_{k-1}), \\ 
\mathbf{V}^{t+1}
&= \textbf{S}_{k}^* \textbf{R}^t + \widehat{\textbf{H}}_k^t,
\label{eq:AMP}
\end{split}
\end{equation}
where we interpret $\textbf{R}^t$ as a {\em residual}, $\widehat{\textbf{H}}_k^{t+1}$ is our current estimate of $\textbf{H}_k$, and
$\mathbf{V}^t$ is the {\em pseudo-data}, which is equal in distribution to $\textbf{H}^{\rm T}_k$ plus i.i.d.\ Gaussian noise with 
variance $\tau_t^2 \approx ||\textbf{R}^t||^2/(M_\mathsf{r}  T_\mathsf{p})$.
Our {\em denoiser}, denoted $\eta(\cdot,\cdot)$, takes as inputs the pseudo-data $\mathbf{V}^t$ and 
SI from the previous block, denoted $\text{SI}_{k-1}$; the form of $\eta(\cdot,\cdot)$ is specified below.
Finally, the residual uses the 
normalized
{\em divergence} of the denoiser,
\begin{equation}
    \label{eq:divergence}
\langle div \,\, \eta_t(\mathbf{V}, \text{SI}) \rangle = \frac{1}{M_\mathsf{r} M_\mathsf{t}} \sum_{i=1}^{M_\mathsf{t}} \sum_{j=1}^{M_\mathsf{r}} \frac{\partial}{\partial \mathbf{V}_{ij}} [\eta_t(\mathbf{V}, \text{SI})]_{ij}.
\end{equation}
We highlight that the conjugate operator is applied elementwise to 
$\textbf{S}_{k}$ when computing $\mathbf{V}^t$ as part of complex AMP~\cite{Maleki_CAMP13}.

\textbf{Candidate Denoisers.} 
Given the distributional properties of the pseudo-data, namely 
$\mathbf{V}^t \approx \textbf{H}_k + \tau_t \mathbf{G}$ where $\mathbf{G}$ has i.i.d. complex Gaussian entries, 
there are at least two plausible denoising styles  for AMP-SI \eqref{eq:AMP}. 

The first denoising style we consider is \emph{conditional expectation},
where $\widehat{\textbf{H}}_k^{t+1}$ is calculated using 
\[
\widehat{\textbf{H}}_k^{t+1} = 
\mathbb{E}[\textbf{H}_k | \mathbf{V}^t = \textbf{H}_k + \tau_t \mathbf{G}, \text{SI}_{k-1} = \widehat{\textbf{H}}_{k-1}],
\]
where we have explicitly stated that the SI at time $k-1$ takes the form 
of an estimate of the channel at the previous block, $\widehat{\textbf{H}}_{k-1}$.  
Within each AMP iteration, conditional expectation provides a {\em minimum mean squared error} (MMSE) 
estimator of $\widehat{\textbf{H}}_k^{t+1}$ given the pseudo-data, $\mathbf{V}^t$, and SI, $\text{SI}_{k-1}$. 
Under some technical conditions, for large scale linear inverse problems~\cite{MontChap11}, 
upon convergence, AMP with conditional expectation denoisers yields the \emph{overall} MMSE signal estimator.  
Unfortunately, with our current understanding of the model, the conditional expectation denoiser appears
computationally intractable when $L_p>1$, so we did not consider it further.

The second denoising style we consider is {\em maximum a posteriori} (MAP),
where we compute the triple
$(\widehat{\theta}_k, \widehat{\Phi}_k, \widehat{\alpha}_k)$ 
that maximizes the posterior,
\begin{equation*}
f(\theta_k, \Phi_k, \alpha_k | \mathbf{V}^t = \textbf{H}_k + \tau_t \mathbf{G}, \text{SI}_{k-1} = (\widehat{\theta}_{k-1}, \widehat{\Phi}_{k-1}, \widehat{\alpha}_{k-1})),
\end{equation*}
where $f(\cdot)$ denotes a generic density, 
and then use the estimated triple $(\widehat{\theta}_k, \widehat{\Phi}_k, \widehat{\alpha}_k)$ 
to produce an estimate of $\widehat{\textbf{H}}_{k}$.
In contrast to the conditional expectation denoiser, 
MAP signal estimation is sub-optimal {in terms of MSE} in individual AMP iterations, 
because conditional expectation is the MMSE estimator, and thus minimizes
the noise variance for the next iteration, \textcolor{black}{whereas} MAP differs from conditional expectation.
Additionally, MAP denoisers may not achieve the overall MMSE.  Despite MAP having these estimation-theoretic drawbacks,
we will see that it offers computational advantages in our analog denoising problem.

For the MAP denoiser, we need to further study the posterior distribution..  First, by Bayes' rule,
\begin{align}
&f(\theta_k, \Phi_k, \alpha_k | \mathbf{V}^t = \textbf{H}_k + \tau_t \mathbf{G},  \text{SI}_{k-1} = (\widehat{\theta}_{k-1}, \widehat{\Phi}_{k-1}, \widehat{\alpha}_{k-1})) \nonumber \\
&= f(\theta_k| \widehat{\theta}_{k-1}) f(\Phi_k| \widehat{\Phi}_{k-1}) f(\alpha_k| \widehat{\alpha}_{k-1}) \times \label{eq:cond_dist2}\\
&\qquad
\frac{ f(\tau_t \mathbf{G} = \mathbf{V}^t -\textbf{H}_k \, | \, \theta_k, \Phi_k, \alpha_k)}{f(\mathbf{V}^t = \textbf{H}_k + \tau_t \mathbf{G} \, | \,  \text{SI}_{k-1} = (\widehat{\theta}_{k-1}, \widehat{\Phi}_{k-1}, \widehat{\alpha}_{k-1}))}, \nonumber  
\end{align}
where we have used the independence of the RVs
$(\theta_k, \Phi_k, \alpha_k)$ and the fact that the pseudo-data at iteration $t$, 
given by $\mathbf{V}^t = \textbf{H}_k + \tau_t \mathbf{G}$, is independent of $(\widehat{\theta}_{k-1}, \widehat{\Phi}_{k-1}, \widehat{\alpha}_{k-1})$ given $(\theta_k, \Phi_k, \alpha_k)$. 
In the numerator (\ref{eq:cond_dist2}), the densities are Gaussian,
\begin{align}
f(\alpha_k | \widehat{\alpha}_{k-1}) &\sim \mathcal{CN}( \rho \widehat{\alpha}_{k-1}, 1{-}\rho^2), \label{eq:individual_dists} \\ 
f(\theta_k| \widehat{\theta}_{k-1})& \sim \mathcal{N}(\widehat{\theta}_{k-1}, \sigma_{\theta}^2), \quad f(\Phi_k| \widehat{\Phi}_{k-1})  \sim \mathcal{N}(\widehat{\phi}_{k-1},  \sigma_{\phi}^2). \nonumber
\end{align}
The denominator is a normalization constant that does not affect MAP optimization.
 
\textbf{MAP Denoiser.} 
Focusing on the MAP denoiser, the form of the conditional distribution given in \eqref{eq:cond_dist2} suggests,
\begin{equation}
\begin{split}
&\argmax_{(\theta_k, \Phi_k, \alpha_k) \in \mathbb{R}^2 \times \mathbb{C}} f(\theta_k, \Phi_k, \alpha_k | \mathbf{V}^t = \textbf{H}_k + \tau_t \mathbf{G}, \text{SI}_{k-1} ) \\
&= \argmax_{(\theta_k, \Phi_k, \alpha_k) \in \mathbb{R}^2 \times \mathbb{C}} \Big\{ \log f(\alpha_k | \widehat{\alpha}_{k-1})  + \log f(\theta_k| \widehat{\theta}_{k-1}) \\
&\hspace{0cm} + \log f(\phi_k| \widehat{\phi}_{k-1}) + \log f(\tau_t \mathbf{G} = \mathbf{V}^t -\textbf{H}_k \, | \,\theta_k, \Phi_k, \alpha_k)\Big\}.
\label{eq:MAP}
\end{split}
\end{equation}
We simplify the above using \eqref{eq:individual_dists}, 
\begin{align*}
&\log f(\alpha_k| \widehat{\alpha}_{k-1}) = - \log(\pi (1 - \rho^2)) - \frac{1}{1 - \rho^2}|\alpha_k - \rho  \widehat{\alpha}_{k-1}|^2, \\
&\log f(\theta_k| \widehat{\theta}_{k-1}) =  -\frac{1}{2} \log(2 \pi \sigma_\theta^2) - \frac{ 1 }{ 2\sigma_\theta^2}(\theta_k - \widehat{\theta}_{k-1})^2,
\end{align*}
\vspace*{-2mm}
\\
where $\log f(\phi_k| \widehat{\phi}_{k-1})$ 
is
similar to $\log f(\theta_k| \widehat{\theta}_{k-1})$, and
\begin{align*}
&\log f(\tau_t \mathbf{G} = \mathbf{V}^t -\textbf{H}_k \, | \, \theta_k, \Phi_k, \alpha_k) \\
&=  \log \prod_{i=1}^{M_{\textsf{r}}} \prod_{j=1}^{M_{\textsf{t}}} f(\tau_t [\mathbf{G}]_{ij} = [\mathbf{V}^t]_{ij} - [\textbf{H}_k]_{ij} \, | \, \theta_k, \Phi_k, \alpha_k) \\
&=  - M_{\textsf{t}} M_{\textsf{r}}\log(\pi)  - \frac{1}{\tau^2_{t}} 
||\mathbf{V}^t - \textbf{H}_k||^2_F,
\end{align*}
where for a matrix $\textbf{M} \in \mathbb{C}^{n \times m}$ we have $\|\textbf{M}\|^2_F = \sum_{i=1}^{n} \sum_{j=1}^{m} |[\mathbf{M}]_{ij}|^2$.
Plugging into \eqref{eq:MAP}, we find
\begin{equation}
\begin{split}
\argmax_{(\theta, \Phi, \alpha) \in \mathbb{R}^2 \times \mathbb{C}} &f(\theta, \Phi, \alpha | \mathbf{V}^t = \textbf{H}_k + \tau_t \mathbf{G}) \\
&= \argmin_{(\theta, \Phi, \alpha_R, \alpha_I) \in \mathbb{R}^4} C_t(\theta, \Phi, \alpha_R, \alpha_I),
\label{eq:MAP_new}
\end{split}
\end{equation}
where
\begin{equation}
\begin{split}
&C_t(\theta, \Phi, \alpha_R, \alpha_I) = \frac{ (\theta - \widehat{\theta}_{k-1})^2 }{ 2\sigma_\theta^2} 
+ \frac{ (\phi - \widehat{\phi}_{k-1})^2 }{ 2 \sigma_\phi^2}\\
&\hspace{.25cm} + \frac{|\alpha_R + j \alpha_I - \rho  \widehat{\alpha}_{k-1}|^2}{1 - \rho^2}  + \frac{1}{\tau^2_{t}} 
||\mathbf{V}^t - \textbf{H}_k||^2_F,
\label{eq:C_func}
\end{split}
\end{equation}
and we have written $\alpha \in \mathbb{C}$ as $\alpha = \alpha_R + j \alpha_I$ with $\alpha_R, \alpha_I \in \mathbb{R}$.  

\textbf{MAP Denoiser Implementation Details.}
Now we discuss the details of implementing the MAP denoiser of \eqref{eq:MAP_new} in the AMP algorithm in \eqref{eq:AMP}, meaning we take 
\begin{align}
&\eta_t(\mathbf{V}^t, \text{SI}_{k-1}) = \argmin_{(\theta, \Phi, \alpha_R, \alpha_I) \in \mathbb{R}^4} C_t(\theta, \Phi, \alpha_R, \alpha_I),
\label{eq:MAP_denoiser}
\end{align}
where we have defined the function $C_t(\cdot)$ in \eqref{eq:C_func}.
In performing the 4D optimization in \eqref{eq:MAP_denoiser}, it is possible to explicitly solve for the 
minimizing pair $(\alpha_R, \alpha_I) \in \mathbb{R}^2$ for any given pair $(\theta, \Phi) \in \mathbb{R}^2$, because
\begin{align*}
&\frac{\partial C_t}{\partial \alpha_R} (\theta, \Phi, \alpha_R, \alpha_I) = \frac{2 (\alpha_R- \rho  
\text{Re}(\widehat{\alpha}_{k-1})) }{1 - \rho^2}  \\
&- \frac{2\kappa}{\tau^2_{t}}\sum_{i=1}^{M_{\textsf{r}}} \sum_{j=1}^{M_{\textsf{t}}} \Big[\text{Re}([\textbf{a}_{\theta,\phi}]_{ij} )\text{Re}([\mathbf{V}^t]_{ij}- \kappa \alpha [\textbf{a}_{\theta,\phi}]_{ij})\\
&\hspace{1cm} -\text{Im}([\textbf{a}_{\theta,\phi}]_{ij} )\text{Im}( [\mathbf{V}^t]_{ij} - \kappa  \alpha [\textbf{a}_{\theta,\phi}]_{ij} )\Big],
\end{align*}
using the shorthand 
$[\textbf{a}_{\theta,\phi}]_{ij} := [\textbf{a}(\theta,M_\mathsf{r})]_i [\textbf{a}(\phi,M_\mathsf{t})^{\rm H}]_j$, $\kappa := \sqrt{ M_\mathsf{t}M_\mathsf{r} \gamma}$, and $\text{{Re}}
(\cdot)$ and $\text{{Im}}(\cdot)$ indicate real and 
imaginary parts, along with the fact that for any $i\in [M_{\textsf{r}}]$ and $j\in [M_{\textsf{r}}]$,
\begin{align*}
&\frac{\partial|[\mathbf{V}^t]_{ij} - [\textbf{H}_k]_{ij}|^2}{\partial \alpha_R}= - 2\kappa \text{Re}([\textbf{a}_{\theta,\phi}]_{ij} )\text{Re}( [\mathbf{V}^t]_{ij}-\kappa\alpha [\textbf{a}_{\theta,\phi}]_{ij})\\
&\hspace{2.5cm} - 2\kappa \text{Im}([\textbf{a}_{\theta,\phi}]_{ij} )\text{Im}( [\mathbf{V}^t]_{ij}-  \kappa \alpha [\textbf{a}_{\theta,\phi}]_{ij} ).
\end{align*} 
For any $(\theta, \phi)$, the minimizing $\widehat{\alpha}_R$ takes the form
\begin{equation*}
\begin{split}
& \widehat{\alpha}_R = \Big[\frac{1}{1 - \rho^2}   + \frac{\kappa^2 }{\tau^2_{t}} \sum_{i=1}^{M_{\textsf{r}}} \sum_{j=1}^{M_{\textsf{t}}} \Big|  [\textbf{a}_{\theta,\phi}]_{ij}\Big|^2\Big]^{-1} \Big[\frac{ \rho  \text{Re}(\widehat{\alpha}_{k-1}) }{1 - \rho^2} + \\
&\frac{\kappa}{\tau^2_{t}} \sum_{i=1}^{M_{\textsf{r}}} \sum_{j=1}^{M_{\textsf{t}}} \text{Re}([\textbf{a}_{\theta,\phi}]_{ij} )\text{Re}([\mathbf{V}^t]_{ij})  + \text{Im}([\textbf{a}_{\theta,\phi}]_{ij} )\text{Im}( [\mathbf{V}^t]_{ij}) \Big].
\end{split}
\end{equation*}
We can similarly show
\begin{equation*}
\begin{split}
& \widehat{\alpha}_I = \Big[\frac{1}{1 - \rho^2}   + \frac{\kappa^2 }{\tau^2_{t} } \sum_{i=1}^{M_{\textsf{r}}} \sum_{j=1}^{M_{\textsf{t}}} \Big|  [\textbf{a}_{\theta,\phi}]_{ij}\Big|^2\Big]^{-1} \Big[\frac{ \rho  \text{Im}(\widehat{\alpha}_{k-1}) }{1 - \rho^2}+   \\
&\frac{\kappa}{\tau^2_{t}} \sum_{i=1}^{M_{\textsf{r}}} \sum_{j=1}^{M_{\textsf{t}}} \text{Re}([\textbf{a}_{\theta,\phi}]_{ij} )\text{Im}( [\mathbf{V}^t]_{ij}) -\text{Im}([\textbf{a}_{\theta,\phi}]_{ij} )\text{Re}([\mathbf{V}^t]_{ij})\Big].
\end{split}
\end{equation*}

In implementing the denoiser in \eqref{eq:MAP_denoiser} within AMP,
for any pair $(\widehat{\phi}, \widehat{\theta})$ we solve for the optimal $\widehat{\alpha}$
using the estimates of $\widehat{\alpha}_R$ and $\widehat{\alpha}_I$ given just above.
Optimal $(\widehat{\phi}, \widehat{\theta})$ are computed using a grid search over values 
within four standard deviations of the SI.
The remaining consideration in implementing \eqref{eq:AMP}
is to compute the divergence of the denoiser \eqref{eq:divergence}. While it is difficult to compute the 
divergence analytically, our implementation in Sec.~\ref{sec:results}
approximates it numerically.

We pause here to note that extending the MAP denoising technique just outlined to two drifting paths, i.e.,\ $L_{\textsf{p}} = 2$ in 
our channel model~\eqref{eq:channel}, seems feasible but computationally expensive, 
and it is unclear if these approaches will work when going beyond, i.e.,\ $L_{\textsf{p}} \geq 3$. 
Extending our results here to these regimes will be pursued in future work.

\section{Numerical Results} \label{sec:results}

\begin{figure}[!t]
\vspace*{-6mm}
\centering
\includegraphics[width=0.5\textwidth]{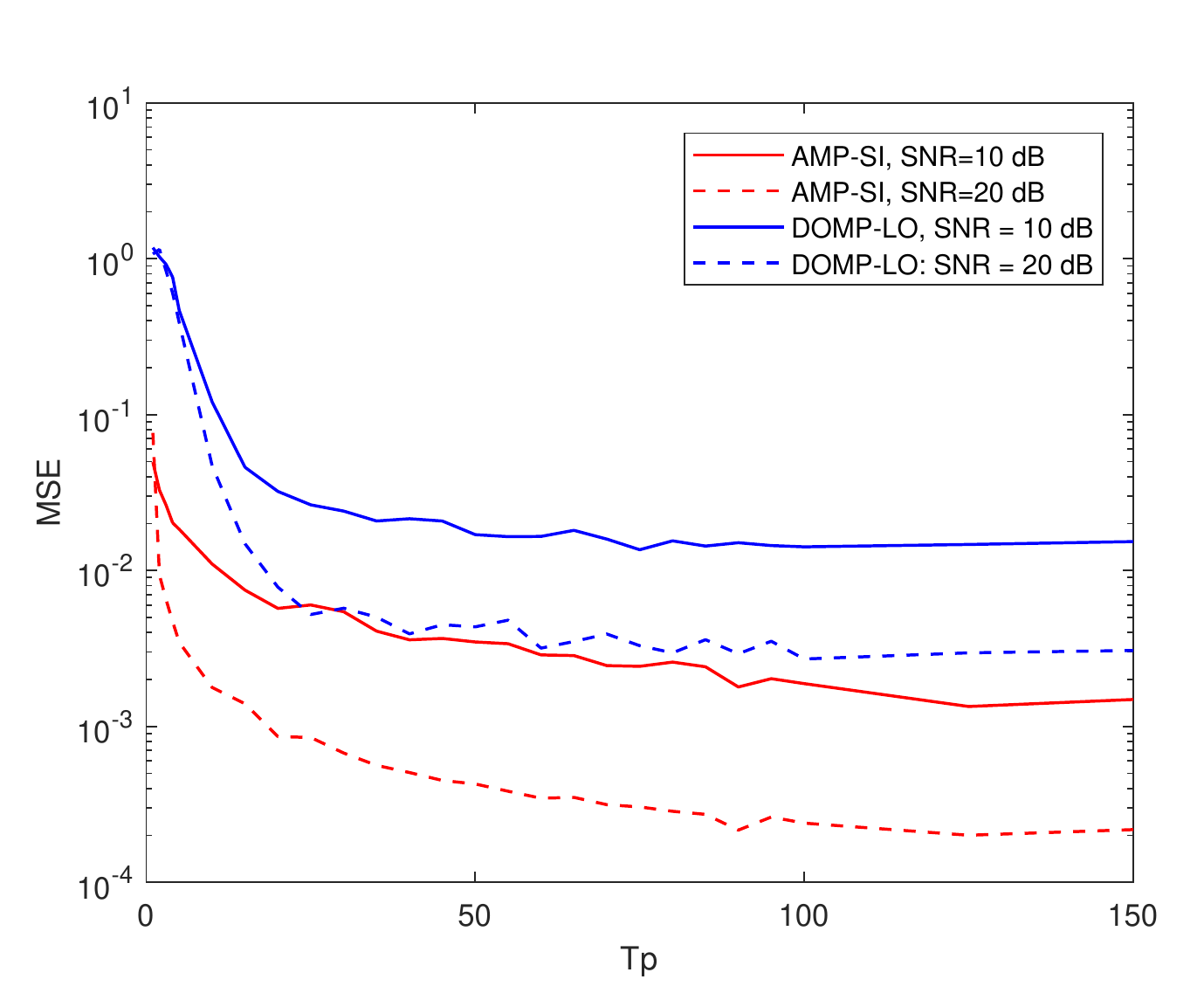}
\vspace*{-8mm}
\caption{MSE as function of number of pilots $T_\mathsf{p}$ per transmission block. (AMP-SI used 300 iterations.)}
\label{fig:exp2_mse_vs_Tp}
\vspace{0.in}
\end{figure}

We demonstrate the performance of the proposed AMP-SI algorithm by presenting numerical results based on Monte Carlo simulations.
As a benchmark, we used {\em Dirichlet orthogonal matching pursuit with local optimization} (DOMP-LO)~\cite{Anjinappa2019ChanEst}, 
which uses Dirichlet kernels while estimating the unknown mmWave channel. Our communications setting used
$M_\mathsf{t} \,{=}\, 16$, $M_\mathsf{r} \,{=}\, 8$, $\rho \,{=}\, 0.995$, $\sigma_\theta^2 \,{=}\, \sigma_\phi^2 \,{=}\, 1$, and $\gamma \,{\in}\, \{10,20\}$ dB, 
which represent a mmWave channel with reasonable time variation. 
Fig.~\ref{fig:exp2_mse_vs_Tp} shows numerical results using 300 AMP-SI iterations;
DOMP-LO used a dictionary size of $32$, which yields the best performance.
It can be seen that the empirical MSE declines as more pilots, $T_\mathsf{p}$, are used.
Moreover, a 20 dB SNR outperformed 10 dB, resulting in lower curves. 
Overall, the empirical MSE obtained by our AMP-SI approach was roughly an order of magnitude
less than that of DOMP-LO.
We also compared our setting to one without SI by increasing the variance of the drift. 
Larger variance reduced the estimation quality; we omit the details for brevity. Finally,
our future work will evaluate the spectral efficiency of the proposed algorithm along with hybrid/digital beamforming schemes in 
comparison to training length, and compare the performance gains over existing methods (e.g., \cite{Cabric2018TraSpa,Love2019MulArm}).

\section{Acknowledgment}
The authors thank Arian Maleki for graciously discussing his work on complex AMP 
\cite{Maleki_CAMP13} with us, and Chethan Anjinappa for help producing numerical results.

\bibliographystyle{IEEEtran}
\bibliography{IEEEabrv,references}

\begin{thebibliography}{10}
\providecommand{\url}[1]{#1}
\csname url@samestyle\endcsname
\providecommand{\newblock}{\relax}
\providecommand{\bibinfo}[2]{#2}
\providecommand{\BIBentrySTDinterwordspacing}{\spaceskip=0pt\relax}
\providecommand{\BIBentryALTinterwordstretchfactor}{4}
\providecommand{\BIBentryALTinterwordspacing}{\spaceskip=\fontdimen2\font plus
\BIBentryALTinterwordstretchfactor\fontdimen3\font minus
  \fontdimen4\font\relax}
\providecommand{\BIBforeignlanguage}[2]{{%
\expandafter\ifx\csname l@#1\endcsname\relax
\typeout{** WARNING: IEEEtran.bst: No hyphenation pattern has been}%
\typeout{** loaded for the language `#1'. Using the pattern for}%
\typeout{** the default language instead.}%
\else
\language=\csname l@#1\endcsname
\fi
#2}}
\providecommand{\BIBdecl}{\relax}
\BIBdecl

\bibitem{DonMalMont09}
D.~Donoho, A.~Maleki, and A.~Montanari, ``Message-passing algorithms for
  compressed sensing,'' \emph{Proc.\ of the National Academy of Sciences}, vol.
  106, no.~45, pp. 18\,914--18\,919, 2009.

\bibitem{krz12}
F.~Krzakala, M.~M{\'e}zard, F.~Sausset, Y.~Sun, and L.~Zdeborov{\'a},
  ``Probabilistic reconstruction in compressed sensing: algorithms, phase
  diagrams, and threshold achieving matrices,'' \emph{Journal of Statistical
  Mechanics: Theory and Experiment}, no.~8, 2012.

\bibitem{MontChap11}
A.~Montanari, ``Graphical models concepts in compressed sensing,'' in
  \emph{Compressed Sensing}, Y.~Eldar and G.~Kutyniok, Eds.\hskip 1em plus
  0.5em minus 0.4em\relax Cambridge University Press, 2012, pp. 394--438.

\bibitem{rangan2019vector}
S.~Rangan, P.~Schniter, and A.~Fletcher, ``Vector approximate message
  passing,'' \emph{IEEE Trans.\ Inf.\ Theory}, 2019.

\bibitem{DonohoCS}
D.~Donoho, ``Compressed sensing,'' \emph{IEEE Trans. Inf. Theory}, vol.~52,
  no.~4, pp. 1289--1306, Apr. 2006.

\bibitem{CandesRUP}
E.~Cand\`{e}s, J.~Romberg, and T.~Tao, ``Robust uncertainty principles: {E}xact
  signal reconstruction from highly incomplete frequency information,''
  \emph{IEEE Trans. Inf. Theory}, vol.~52, no.~2, pp. 489--509, Feb. 2006.

\bibitem{Ma2018}
A.~Ma, Y.~Zhou, C.~Rush, D.~Baron, and D.~Needell, ``An approximate message
  passing framework for side information,'' \emph{IEEE Trans. Signal Proc.},
  vol.~67, no.~7, pp. 1875--1888, Apr. 2019.

\bibitem{Liu2019ISIT}
H.~Liu, C.~Rush, and D.~Baron, ``An analysis of state evolution for approximate
  message passing with side information,'' in \emph{Proc. IEEE Int. Symp. Inf.
  Theory}, July 2019.

\bibitem{Rappaport2016mmW}
M.~{Samimi} and T.~{Rappaport}, ``3-{D} millimeter-wave statistical channel
  model for {5G} wireless system design,'' \emph{IEEE Trans. Microw. Theory
  Techn.}, vol.~64, no.~7, pp. 2207--2225, Jul. 2016.

\bibitem{Pati1993}
Y.~C. Pati, R.~Rezaiifar, and P.~S. Krishnaprasad, ``{Orthogonal matching
  pursuit: {R}ecursive function approximation with applications to wavelet
  decomposition},'' \emph{Proc. 27th Asilomar Conf. Signals, Syst. Comput.},
  pp. 40--44, Nov. 1993.

\bibitem{Guo2016TraAng}
C.~Zhang, D.~Guo, and P.~Fan, ``Tracking angles of departure and arrival in a
  mobile millimeter wave channel,'' in \emph{Proc. IEEE Int. Conf. Commun.
  (ICC)}, May 2016, pp. 1--6.

\bibitem{Heath2016BeaTra}
V.~Va, H.~Vikalo, and R.~Heath, ``Beam tracking for mobile millimeter wave
  communication systems,'' in \emph{IEEE Global Conf. Signal Inf. Process.
  (GlobalSIP)}, Dec. 2016, pp. 743--747.

\bibitem{Kim2017RobBea}
S.~Jayaprakasam, X.~Ma, J.~Choi, and S.~Kim, ``Robust beam-tracking for
  {mmWave} mobile communications,'' \emph{IEEE Commun. Lett.}, vol.~21, no.~12,
  pp. 2654--2657, Dec. 2017.

\bibitem{Tan_CompressiveImage2014}
J.~Tan, Y.~Ma, and D.~Baron, ``Compressive imaging via approximate message
  passing with image denoising,'' \emph{IEEE Trans. Signal Process.}, vol.~63,
  no.~8, pp. 2085--2092, Apr. 2015.

\bibitem{Metzler2016_IEEE}
C.~Metzler, A.~Maleki, and R.~Baraniuk, ``From denoising to compressed
  sensing,'' \emph{IEEE Trans. Inf. Theory}, vol.~62, no.~9, pp. 5117--5144,
  Sept. 2016.

\bibitem{Ma_nonseparable19}
Y.~Ma, C.~Rush, and D.~Baron, ``Analysis of approximate message passing with
  non-separable denoisers and {Markov} random field priors,'' \emph{IEEE Trans.
  Inf. Theory}, vol.~65, no.~11, pp. 7367--7389, Nov. 2019.

\bibitem{Hamzehei2015compressive}
S.~Hamzehei and M.~Duarte, ``Compressive parameter estimation via approximate
  message passing,'' in \emph{2015 IEEE Int.\ Conf.\ on Acoustics, Speech and
  Signal Process.}\hskip 1em plus 0.5em minus 0.4em\relax IEEE, 2015, pp.
  3327--3331.

\bibitem{Hamzehei2016compressive}
------, ``Compressive direction-of-arrival estimation off the grid,'' in
  \emph{2016 50th Asilomar Conference on Signals, Systems and Computers}.\hskip
  1em plus 0.5em minus 0.4em\relax IEEE, 2016, pp. 1081--1085.

\bibitem{Bellili2019generalized}
F.~Bellili, F.~Sohrabi, and W.~Yu, ``Generalized approximate message passing
  for massive {MIMO} mmwave channel estimation with {Laplacian} prior,''
  \emph{IEEE Trans.\ on Comm.}, vol.~67, no.~5, pp. 3205--3219, 2019.

\bibitem{Maleki_CAMP13}
A.~Maleki, L.~Anitori, Z.~Yang, and R.~Baraniuk, ``Asymptotic analysis of
  complex lasso via complex approximate message passing (camp),'' \emph{IEEE
  Trans. Inf. Theory}, vol.~59, no.~7, pp. 4290--4308, July 2013.

\bibitem{Anjinappa2019ChanEst}
C.~K. Anjinappa, Y.~Zhou, Y.~Yap{\i}c{\i}, D.~Baron, and I.~G\"{u}ven\c{c},
  ``Channel estimation in {mmWave} hybrid {MIMO} system via off-grid
  {Dirichlet} kernels,'' in \emph{Proc. {IEEE} Global Commun. Conf.
  (GLOBECOM)}, Waikoloa, Hawaii, Dec. 2019.

\bibitem{Cabric2018TraSpa}
V.~{Boljanovic}, H.~{Yan}, and D.~{Cabric}, ``Tracking sparse {mmWave} channel
  under time varying multipath scatterers,'' in \emph{Proc. Asilomar Conf.
  Signals Syst. Comput.}, 2018, pp. 1274--1279.

\bibitem{Love2019MulArm}
M.~B. {Booth}, V.~{Suresh}, N.~{Michelusi}, and D.~J. {Love}, ``Multi-armed
  bandit beam alignment and tracking for mobile millimeter wave
  communications,'' \emph{IEEE Commun. Lett.}, vol.~23, no.~7, pp. 1244--1248,
  2019.

\end{thebibliography}

\end{document}